# Using metadynamics to build neural network potentials for reactive events: the case of urea decomposition in water

Manyi Yang [1,3], Luigi Bonati [2,3], Daniela Polino [1,3,*] and Michele Parrinello [1,3,4,**]

1 Department of Chemistry and Applied Biosciences, ETH Zurich, 8092 Zurich, Switzerland
2 Department of Physics, ETH Zurich, 8092 Zurich, Switzerland
3 Institute of Computational Sciences, Università della Svizzera italiana, Lugano, Switzerland
4 Italian Institute of Technology, Via Melen 83, 16152 Genova, Italy
* *daniela.polino@usi.ch* , ***michele.parrinello@phys.chem.ethz.ch*

## Abstract

The study of chemical reactions in aqueous media is very important for its implications in several fields of science, from biology to industrial processes. However, modeling these reactions is difficult when water directly participates in the reaction, since it requires a fully quantum mechanical description of the system. *Ab-initio* molecular dynamics is the ideal candidate to shed light on these processes. However, its scope is limited by a high computational cost. A popular alternative is to perform molecular dynamics simulations powered by machine learning potentials, trained on an extensive set of quantum mechanical calculations. Doing so reliably for reactive processes is difficult because it requires including very many intermediate and transition state configurations. In this study we used an active learning procedure accelerated by enhanced sampling to harvest such structures and to build a neural-network potential to study the urea decomposition process in water. This allowed us to obtain the free energy profiles of this important reaction in a wide range of temperatures, to discover several novel metastable states, and improve the accuracy of the kinetic rates calculations. Furthermore, we found that the formation of the zwitterionic intermediate has the same probability of occurring via an acidic or a basic pathway, which could be the cause of the insensitivity of reaction rates to the solution pH.

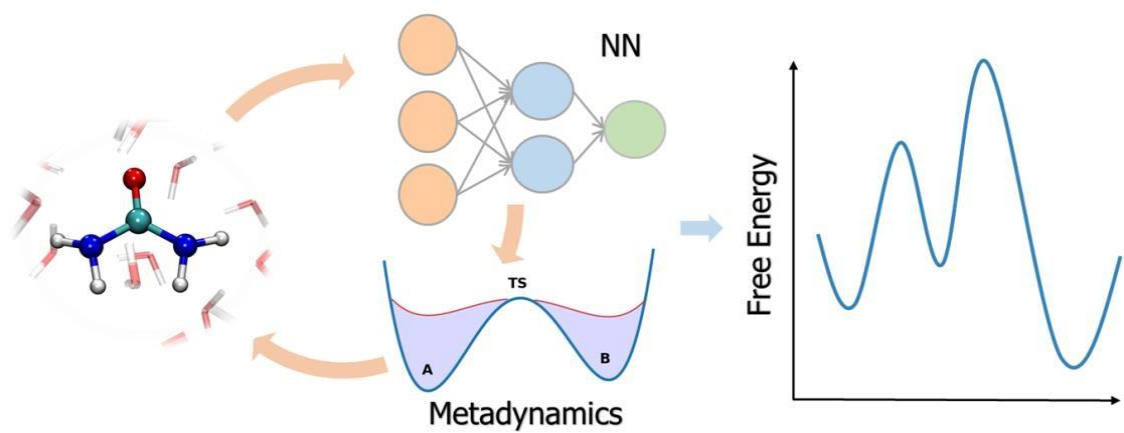

# 1. Introduction

Liquid phase reactions are routinely used in the laboratory practice and play an important role in chemistry, biology and industrial processes. In order to study this important class of reactions several theoretical tools have been developed. Since computing the properties of the whole system is rather expensive, the solvent has been often approximated as a continuum dielectric[1–3] or treated in a quantum-mechanics/molecular mechanics (QM/MM) scheme[4,5]. However, it is possible to remove these approximations and treat the whole system at the quantum level in what is known as *ab-initio* molecular dynamics (AIMD). This approach, although costly, has several advantages. It treats solvent and reactants at the same level, and it allows any of the molecules of the solvent to be part of the reaction and samples the solvent fluctuations.

These AIMD features are of particular relevance in the study of reactions in water, in which proton transfers are often assisted by solvent molecules[6] and the fluctuations of the hydrogen bonded network of water play a significant role. In a recent AIMD study of the decomposition of urea in water, we have shown that these two features were highly relevant[7]. However, AIMD does not come cheaply and this has limited the thoroughness of the sampling.

In this paper, we show that a properly designed strategy of active learning allows constructing a reactive potential energy surface of *ab-initio* quality that can be used to improve considerably sampling and to extend the scope of AIMD simulations.

We illustrate our strategy in the case of urea decomposition in water. We have chosen this example for its relevance to biology[8], agriculture[9], medical technology[10], energy[11,12] and environment[13,14]. Given its importance, it is not surprising that over the past century much experimental[15–19] and theoretical effort[7,20–22] has been devoted to its study. Experiments suggest that the reaction proceeds via the formation of a Zwitterionic intermediate ($^+NH_3CONH^-$) [17] and a successive $NH_3$ elimination, following the kinetic scheme:

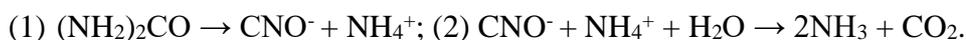

(1) $(NH_2)_2CO \rightarrow CNO^- + NH_4^+$; (2) $CNO^- + NH_4^+ + H_2O \rightarrow 2NH_3 + CO_2$.

However, there is some uncertainty as to the value of the free energy barrier that separates reactants from products with estimates that range from 119 to 136 kJ/mol[15]. In addition, also some aspects of the reaction are not clear like for instance the reaction insensitivity to pH[23,24].

Jorgensen and collaborators have performed a QM/MM calculation in which only one of the water molecules has been treated at the QM level[21]. Their results are in agreement with the scheme reported above. More recently our group has performed an AIMD simulation of the same system[7]. Our result confirmed once again the accepted mechanism, but a more complex picture did emerge in which water and water fluctuations are an important and complex part of the reaction coordinate. However computational cost prevented us from fully converging the free energy of the system and explore the properties of the system at lower temperatures.

Here we removed these limitations, while at the same time retaining the AIMD accuracy. This result has been obtained following the strategy pioneered by Behler and Parrinello [25] which made use of a deep neural network (NN) to represent the potential energy surface of the system. Lately, this approach has become very popular [26–35] and also new methods[36–39] and codes have been made available making the job of training a NN potential easier.

In order for this procedure to be successful, the NN has to be presented with a large number of accurate DFT calculations, performed on an appropriate set of configurations. The choice of the configurations is of the outmost importance, and it is particularly relevant here in a reactive multicomponent system where the NN has to be able to predict the energy of configurations with very different chemical structures. Besides, since reactions proceed via rarely visited transition states it is imperative that such configurations are included in the training set. Given the height of the barriers characteristic of this system, transition state configurations can be accessed only using enhanced sampling techniques such as metadynamics[40,41]. Thus, we shall extensively use metadynamics that will be an integral part of the construction of the NN potential, as done already in our recent works on elemental systems[26,30]. In such a manner we could significantly extend our previous AIMD calculations[7] exploring much lower temperatures, converging the free energies and discovering a number of novel metastable states, and identifying that the zwitterionic intermediate state could be formed either via an acidic or a basic reaction pathway. Furthermore, we were able to significantly improve the accuracy with which the kinetic parameters are calculated.

## 2. Theoretical Methods

### 2.1 Enhancing configurational sampling with metadynamics

Metadynamics[40,41] is a well-established enhanced sampling method in which a history-dependent bias potential $V(s)$ is constructed as a function of atomic coordinates $R$ via a restricted number of collective variables (CVs), $s = s(R)$. By gradually enhancing the fluctuations on the CVs, large energy barriers can be overcome so that rare events are accelerated and take place in an affordable computational time. Eventually, the statistical properties of the system such as the free energy surface (FES) can be computed via a reweighting procedure. In the present work, all enhanced sampling simulations are performed with the Well-Tempered version[42] of metadynamics (WTMetaD), in which a Gaussian $G(s, s_k)$ centered at the visited point $s_k$ is added periodically to the potential as described by:

$$V_n(s) = \sum_{k=1}^{n} e^{-\beta/(\gamma-1)V_{k-1}(s_k)} G(s, s_k) \qquad (1)$$

where $\beta = 1/k_B T$ is the inverse temperature, the parameter $\gamma > 1$ is called the bias factor, and $G(s, s_k) = W e^{\frac{-\|s-s_k\|^2}{2\sigma^2}}$ is a Gaussian function centered at the instantaneous position $s_k$, with $W$ and $\sigma$ being the height and width of the Gaussian, respectively.

Following Ref. [43] and our previous work[7] we resorted to path collective variables. These measures the progression and the distance relative to a preassigned reaction path. The path is defined on the space defined by a set of descriptors which we represent with a *L*-dimensional vector **d**. In this space the path is discretized and is defined by *M* reference nodes $\mathbf{d}^i \forall i = 1, 2, \cdots, M$. In the practice the path CVs are written as:

$$s(\mathbf{d}) = \frac{\sum_{i=1}^{M} i e^{-\lambda \|\mathbf{d}-\mathbf{d}^i\|^2}}{\sum_{i=1}^{M} e^{-\lambda \|\mathbf{d}-\mathbf{d}^i\|^2}} \quad (2)$$

$$z(\mathbf{d}) = -\frac{1}{\lambda} \log[\sum_{i=1}^{M} e^{-\lambda \|\mathbf{d}-\mathbf{d}^i\|^2}] \quad (3)$$

In this work, we used as descriptors **d** the set of coordination numbers: $C_{CN}$, $C_{CO}$ and $C_{N1H} - C_{N2H}$. A detailed description of the CVs adopted in this work is given in the Supporting Information (SI).

**2.2 Kinetic rates estimation via infrequent metadynamics**

We adopted the infrequent metadynamics protocol[44] to calculate the kinetic rates. Infrequent metadynamics is a modified version of metadynamics specifically designed to estimate the kinetic rates of rare events. This method assumes that if no bias is deposited on the transition region then the rapid well-to-well dynamics remains unaffected. In the practice, this assumption can be satisfied by using an infrequent deposition of small Gaussians. However, due to the high free energy barrier of urea decomposition, this would require several long-time simulations to obtain accurate characteristic times. To accelerate the calculations, we have added an external static potential that filled the reactants basin. In order to compare our results with those obtained in Ref. [7], we adopted the same strategy to build this static bias. A more detailed description is reported in the SI. This bias was then used as an external potential, and on this biased energy surface, an infrequent metadynamics run was performed. The physical transition times $t^*$ can be then recovered as:

$$t^* = t_{MD} \left\langle e^{\beta(V(s)_{MetaD} + V(s)_{static})} \right\rangle_V \quad (4)$$

where $t_{MD}$ is the MD simulation time and the average $\left\langle e^{\beta(V(s)_{MetaD} + V(s)_{static})} \right\rangle_V$ is the acceleration factor that measures how much the escape time is boosted because of the presence of the static bias $V(s)_{static}$ and the infrequent metadynamics one $V(s)_{metad}$. The characteristic time can be inferred from a fit of the transition times to a Poisson distribution, and its accuracy assessed by a statistical analysis based on the Kolmogorov-Smirnov (KS) test. We refer the interested reader to Ref.[45] for further details.

## 2.3 Neural network representation of the potential energy surface

Our machine learning-based potential has been constructed following the Deep Potential MD (DPMD) scheme recently developed by Zhang *et al.*[38,46], which has been shown to accurately reproduce the interatomic forces and energies predicted by ab initio calculations in several condensed matter systems. Like in the Behler-Parrinello approach[25], a deep neural network architecture is used to represent the PES, which is decomposed in the sum of atomic energies $E_i$, depending on the local environments, i.e., $E = \sum E_i$. In order to preserve the PES symmetries, a local embedding network [47] is used to project the coordinates of the neighbors of each atom within a cutoff radius $R_c$ into descriptors, which are then used as inputs for the fitting NN. This allows dealing more efficiently with multiple chemical species, without the need of introducing ad-hoc symmetry functions. Hereafter, DPMD-based molecular dynamics simulations and their variants accelerated by metadynamics will be referred to as DPMD and DP-WTMetaD, respectively.

## 2.4 Active-learning optimization accelerated by enhanced sampling

The collection of the reference configurations used in the optimization of the NN is a crucial step that directly affects the validity of the resulting potential. Although NNs are good at interpolating between training points, they cannot predict the energies and forces of configurations that are distant from those of the training set. For this reason, it is critical to include all the configurations relevant to the process at the thermodynamic conditions of interest. This task is far from trivial, especially in the case of activated processes. In fact, one needs to explore a vast chemical space that includes reactants, products, and intermediate states along all the possible reaction pathways.

To address this issue, we take inspiration from our previous works in which we have used enhanced sampling methods to collect training configurations in the study of silicon crystallization[26] and gallium phase diagram[30]. We use the same strategy exploited in the study of gallium, as described below. The NN training is done on configurations harvested at different temperatures using an active learning procedure accelerated by metadynamics simulations. This procedure is schematically illustrated in Fig. 1 and goes as follows.

First, a NN potential is trained on a small set of structures randomly extracted from an AIMD WTMetaD reactive trajectory. This initial NN potential is regarded as the starting point of an iterative training procedure consisting of the following steps:

(i) exploration of the configurational space via DP-WTMetaD simulations at multiple temperatures,

(ii) selection of a small set of relevant configurations to be included in the training set,

(iii) calculation of energies and atomic forces at the DFT level,

(iv) training an ensemble of NN potentials with the updated training set.

The criterion used to label the new configurations in step (ii) is the standard deviation on atomic forces over an ensemble of NN potentials, initialized with different weights[48,49]. Whenever the maximum over the atoms exceeds a predefined threshold, the configuration is selected, and its energy and forces are computed in step (iii). This iterative procedure continues until the number of newly labeled configurations becomes smaller than a preassigned number $N_{min}$. Additional details on the NN training can be found in the SI.

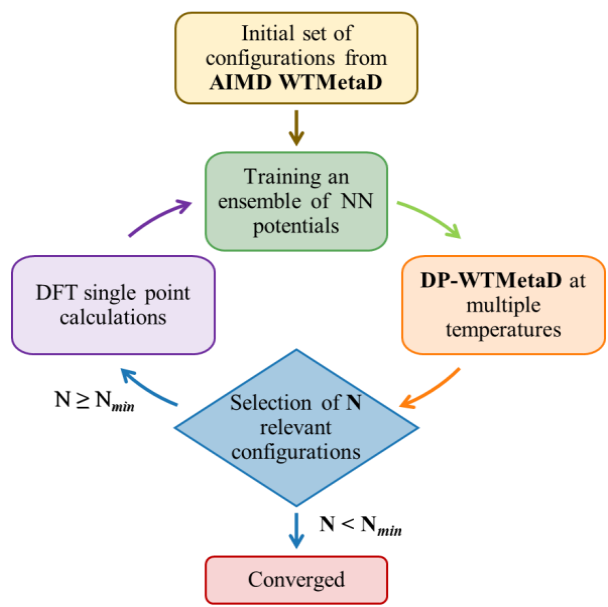

Fig. 1. *Flowchart of the active learning iterative procedure used to train the neural network potential.*

However, labeling the configurations based on the the forces deviation did not provide a uniform coverage of the whole reaction pathway. Indeed, we found that less than 5% of the structures did belong to the transition state region (see Table S1), leading to higher errors on the energies of these configurations (see Fig. S3(a)). This is related to the nature of well-tempered metadynamics, whose aim is to lower the free energy barriers, not to cancel them, so as to prevent the system from exploring unphysically high free energy regions. As a consequence, most of the simulation time is spent within the metastable states, while the time in which the system passes through the transition state is very short.

Since we aim at computing kinetic rates, which depend exponentially on transition state energies, we need to model the transition state configurations as accurately as possible. To address the imbalance between the number of reactants and products configurations versus the number of those belonging to the transition state, we included extra structures generated by the final NN potential from long DP-WTMetaD simulations. The samples are extracted along a CV which is correlated with the reaction pathway. This can be seen as using a different criterion for the active learning procedure, namely to select new configurations with a uniform coverage of the reaction process. Indeed, this resulted in a uniform error along the reaction pathway, as reported in Fig. S3(b). The non-uniform sampling problem could be addressed also by resorting to different enhanced sampling methods to accelerate the active learning procedure. An interesting candidate for chemical reactions could be a recently developed method which focuses the sampling into the transition state[50].

It is worth noticing that the combination of WTMetaD and an active learning procedure[48] enables us to build a reliable NN potential with minimal human intervention and reduced computational cost. In our final training set, only around 10% of the configurations came from the initial AIMD trajectory at 490 K, while the others are collected automatically via inexpensive DP-WTMetaD simulations at multiple temperatures (300, 390, and 490 K). Indeed, in our case DP-WTMetaD calculations are about 500 times faster than AIMD for similar system sizes. For larger systems the cubic scaling of AIMD would make this comparison even more favorable.

This extended strategy allowed us to incorporate all the thermodynamically accessible structures in the training set, covering a wider portion of the configurational space when compared to AIMD WTMetaD simulations (Fig. 2). Notably, we were able to locate some new intermediate states and pathways, which could not be explored via AIMD simulations due to their infrequent character. Including such a variety of relevant configurations in the training set is key to train an *ab-initio* quality NN potential which enables us to study the mechanisms and dynamics of a complex reactive event like urea decomposition in water.

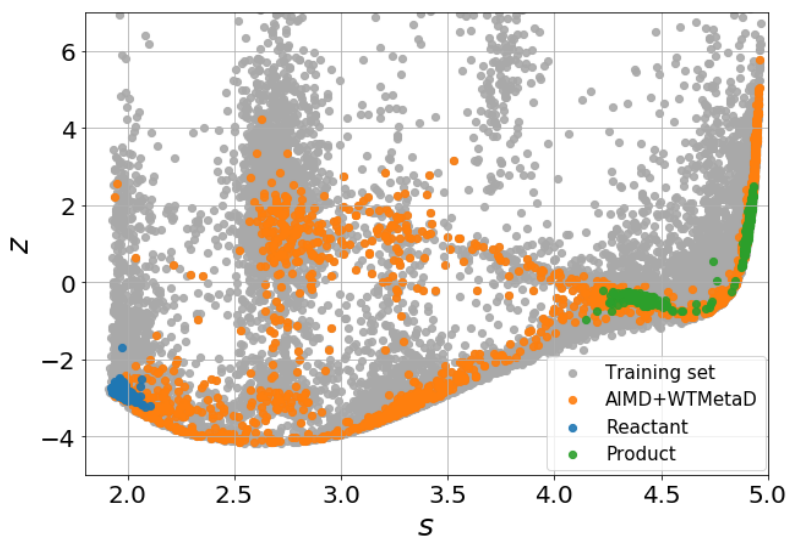

Fig. 2. *Training set configurations in the path collective variables space. The* s *variable represents the progress along the path between reactants and products, while* z *represents the distance from the reference path. Both quantities are unitless. Grey points correspond to the training set configurations sampled at multiple temperatures (300, 390, and 490 K). The orange points correspond to the configurations extracted from the WTMetaD AIMD trajectory at 490 K. For comparison, structures of reactants and products from the AIMD trajectory are also presented with blue and green points, respectively.*

**2.5 Validation of MetaD-based NN Potential**

We first checked the ability of the MetaD-based NN potential to reproduce DFT results in the temperature range from 300 to 500 K. The mean absolute errors (MAEs) on the training set are equal to 7.00 kJ/mol for the energies and 3.71 kJ/mol/Å for the forces. The MAEs on a test set composed by configurations generated by the final DPMD potential are equal to 6.05 kJ/mol for the energies and 3.54 kJ/mol/Å for the forces (see also Fig. S1-S3 for the MAE as a function of a collective variable). It is remarkable that, for each temperature (300, 390, and 490 K), the distribution of the energy MAEs along the reaction pathway is uniform (see Fig. S3). That implies that our NN potential can describe the configurations from the whole FES with uniform accuracy.

Furthermore, we compared the equilibrium properties of urea in aqueous solution sampled by DPMD and AIMD at five different temperatures from 300 to 500 K. Both the conformational distributions (Fig. S4) and the radial distribution functions (Figs. S5 to S9) computed via DPMD agree with the AIMD results. These results suggest that the configurational space of urea in aqueous media is well reproduced within the wide temperature range of 300 ~ 500 K.

## 3. Results and discussion

**3.1 Free energy surface unveils urea decomposition mechanism**

We used the DPMD potential determined as described above to obtain converged free energy surfaces (FES) at T = 350, 390, 450, and 490 K. We checked the convergence of the calculations by analyzing the time evolution of the free energy difference between reactants and products ($\Delta F_{AB}$), which is reported in Fig. S10. The standard deviation on $\Delta F_{AB}$ calculated from 4 independent simulations is less than 2 kJ/mol. Here we describe in detail the T = 350 K surface that is representative of the system behavior. The FES corresponding to higher temperatures are reported in Fig. S11, while a short discussion of the FES temperature dependence is given below.

The FES landscape can be described as composed of two complex basins. The solvated urea reactant state (A1) is at the bottom of basin A. In the same basin one finds two other metastable states A2 and A3. In the metastable state A2 the neutral molecule isourea ($HNCOHNH_2$) is present while in A3 one finds the zwitterionic state, $^-HNCONH_3^+$. State A2 does not play a role, while for the reaction to take place the system has first to go through the zwitterionic state and then pass via a high energy transition state in which the C-$NH_3^+$ bond is broken. This leads to basin B where one finds the product state B1, $NCO^-$ + $NH_4^+$ and the short lived intermediate B2, $HNCO$ + $NH_3$.

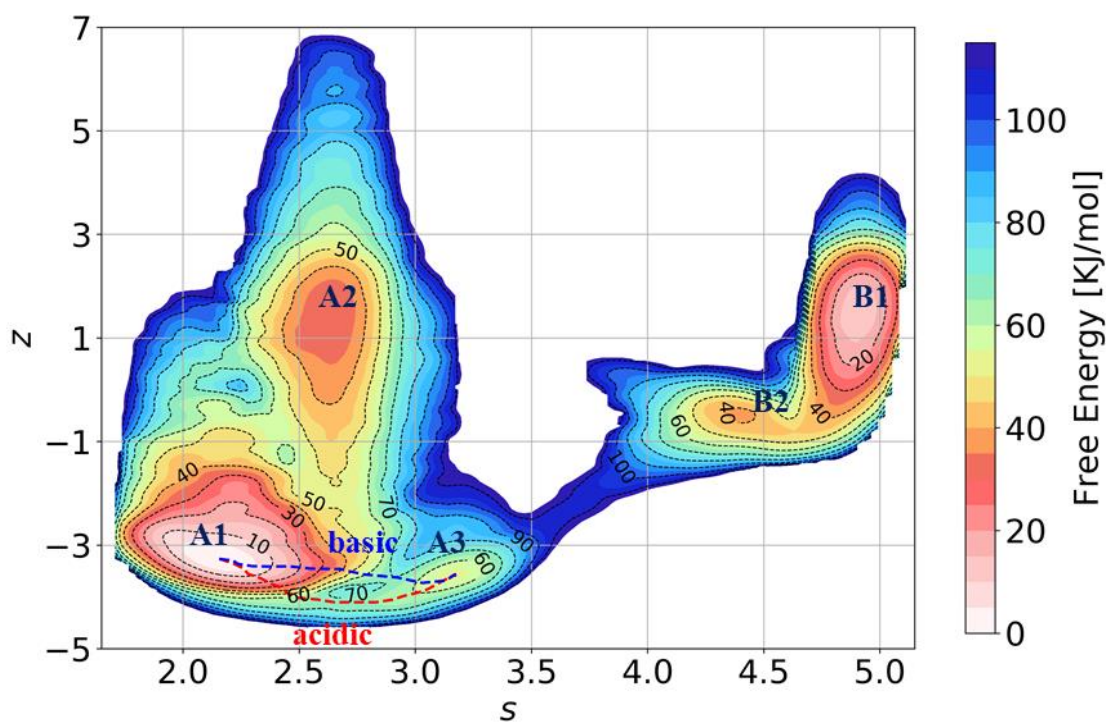

Fig. 3. *Free energy surface of urea decomposition in aqueous solution at 350 K as a function of path collective variables s and z. The dashed lines correspond to the minimum free-energy pathways between A1 and A3, which have been computed using the nudged elastic band method.*

Intra-basins transitions take place via water assisted proton transfers. We illustrate here two typical examples that are relevant to the crucial A1 to A3 transition, as shown in Fig. 4. Two different kinds of reactive events are possible. In the first one, that we refer to as acidic, one water molecule donates a proton to an aminic group, leaving behind a hydroxide $OH^-$ anion. If the $OH^-$ is properly placed, it can take a proton out from the other aminic group, forming the urea zwitterion in a concerted mechanism. Otherwise, the hydroxide can diffuse via a Grotthus mechanism and then eventually reach the aminic residues completing the zwitterionic structure. In the second type of reaction pathway, that we call basic, water acts as a base and accepts a proton from an aminic group leaving behind an hydronium cation, that is transferred via the water network to the other amine using a similar mechanism[51]. If we compare these two transition pathways, we can see from Fig. 5 that they have approximately the same probability of taking place. As discussed in our previous work on urea[7], there is not a unique way in which water ions can move from one amine to another since their transfer can be mediated by different water arrangements. Thus, one cannot identify a single transition state, but one must talk, as done in the path sampling literature[52] of an ensemble of transition states. We also note that, besides being instrumental in allowing the proton transfer that leads to the zwitterionic structure, water helps in stabilizing the structure by screening the two zwitterionic charges.

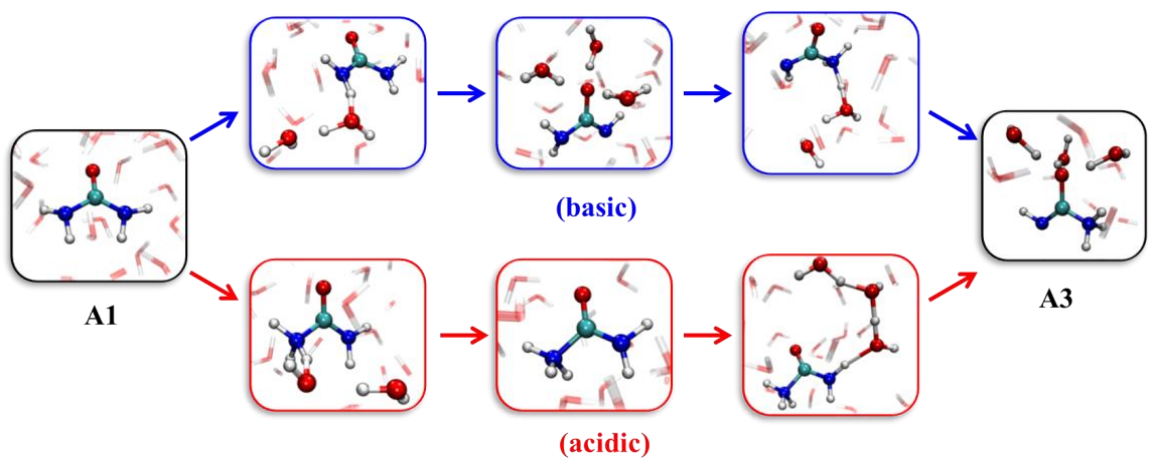

Fig. 4. *Typical snapshots along the acidic (bottom red) and basic (top blue) pathways for the intra-basins transition between reactant A1 and zwitterion intermediate A3. For simplicity, only the solute and few water molecules that are directly involved in the reaction center are shown in ball-and-stick type.*

While for intra-basins transitions it is not possible to identify a unique transition state, the transition from A to B is clearly defined by the breaking of the C-$NH_3^+$ bond. We have checked this to be the case performing a committor analysis on the reactive trajectories [53–55]. The configurations for which the systems can go with equal probability to reactants or to products are characterized by well-defined C-$NH_3^+$ and C-$N^-$ distance (see Table 1). We have also calculated the critical value at which the C-$NH_3^+$ breaks using a standard implicit solvent static calculation (see SI for details). The fact that the two estimates are close is a vindication of the quality of our NN potential in the TS region.

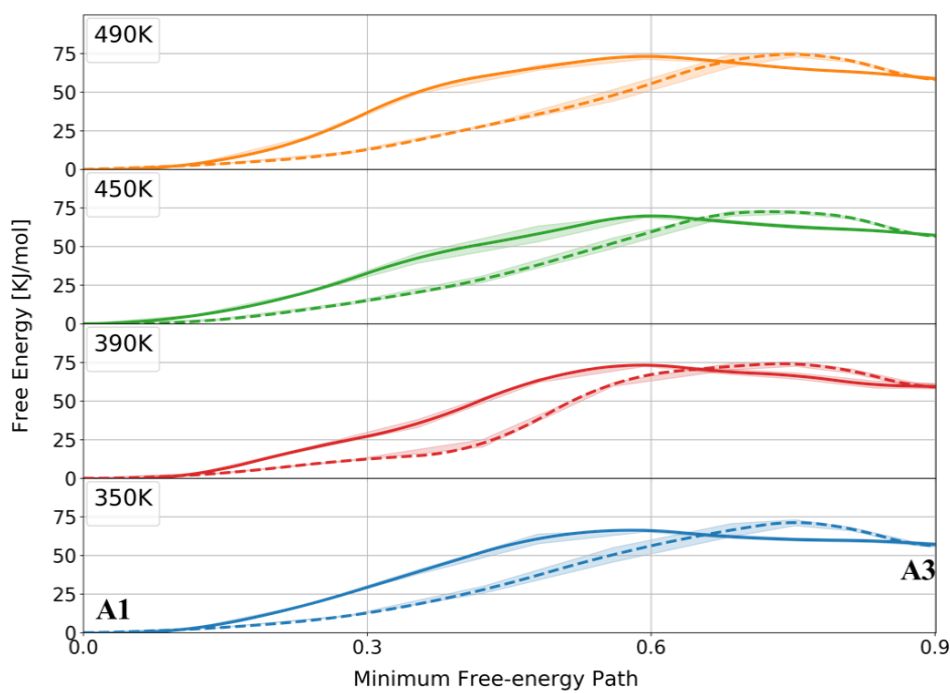

Fig. 5. *Comparison of the acid (solid line) and base (dashed line) catalyzed minimum free-energy pathways for the intra-basins transition at different temperatures. Shaded areas indicate the standard errors obtained from 4 independent DP-WTMeatD simulations.*

The breaking of the C-NH$_3^+$ bond is accompanied by an internal electronic rearrangement and leads first to the two neutral moieties HNCO and NH$_3$ (B1). These are very short-lived and once again a water mediated proton transfer takes place leading to the two oppositely charged moieties NCO$^-$ and NH$_4^+$ (B2) that, due to water screening, can quickly separate.

Table 1: *Bond distances of C-NH3$^+$ and C-N$^-$ in the transition state ensemble between A3 and B2. For each distance we reported the average value and its standard deviation.*

| T | $d$C-NH$_3^+$[Å] | $d$C-N$^-$[Å] |
|---|---|---|
| 350 K | 2.16 ± 0.09 | 1.26 ± 0.02 |
| 490K | 2.18 ± 0.07 | 1.25 ± 0.03 |
| PCM[a] | 2.10 | 1.27 |

[a]PCM: Polarizable Continuum Model[2] at the PBE/cc-pVDZ level[56]

Increasing the temperature of the system (T > 350 K), the main features of the free energy surface described above remain unaltered, however a few quantitative changes can be observed. The difference in free energy $\Delta F_{AB}$ depends approximately linearly on the temperature (see Fig. 6). From the dependence of $\Delta F_{AB}$ of T the change in entropy during the reaction $\Delta S_{AB}$ can be computed from the relation $\Delta S_{AB} = -(\frac{\partial \Delta F_{AB}}{\partial T})_v$, in this way we estimate $\Delta S_{AB}$ to be 28.3 ± 2.1 J/mol/K.

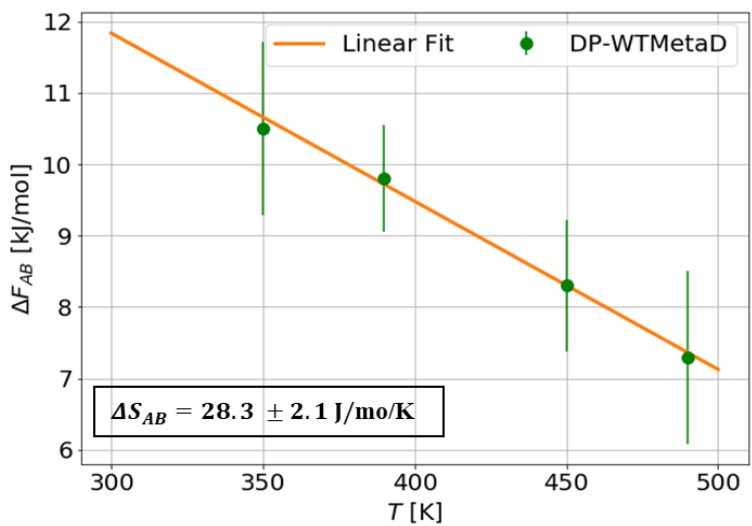

Fig. 6. *Free energy difference $\Delta F_{AB}$ from reactant (A1) to product (B1) at different temperatures and the linear fit of the $\Delta F_{AB}$ as the function of temperature. The error bars on $\Delta F_{AB}$ were calculated from the standard deviation of four independent DP-WTMetaD simulations, and the final entropy error estimate (± 2.1 kJ/mol) is obtained from the uncertainty of the linear fit.*

In closing this section, we note that the speed-up provided by the NN potential gave us the possibility of exploring a larger portion of the configurational space along the reaction pathway, and to observe all the thermodynamically accessible pathways and intermediates, measuring quantitatively their relative probability. Notably, in the AIMD study [7] only one pathway was detected for the formation of the zwitterion ($^+NH_3CONH^-$), while in the present work we observed two reaction mechanisms which are equally possible. Therefore, even if acidic or basic conditions would impede one of the two channels, the formation of the zwitterion could still be possible via the other one. We make the hypothesis that this could be one of the reasons why reaction rates are insensitive to the solution pH.

### 3.2 Kinetic rates calculations

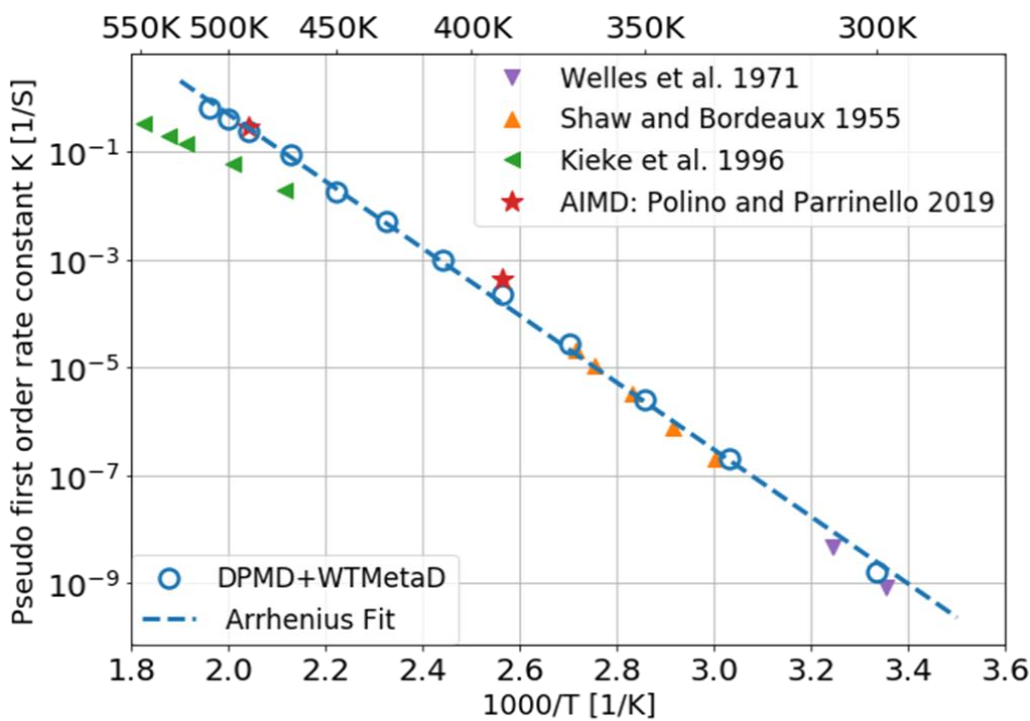

Fig. 7. *Pseudo-first-order rate constants for urea decomposition in aqueous solution as a function of temperature calculated with WTMetaD-based NN potential (blue circles), and comparison with AIMD (red stars) as well as different experimental results (triangles). The least-squares Arrhenius fit is displayed as the blue dotted line.*

Using infrequent metadynamics as described earlier we were able to harvest for each of the 12 temperatures investigated hundreds of reactive trajectories ($A1 \rightarrow B2$). At each temperature we analyzed the distribution of exit times and observed that they are close to being Poisson-like (Fig. S12 and Table S2). From this analysis we extracted the reaction rates which we reported in the Arrhenius plot in Fig. 7. Whenever comparison was possible these results are close to our previous ab-initio simulation estimates. They are also close to the experimental results. However some more marked differences can be seen with a set of high temperature results[19]. The theoretical Arrhenius plot is slightly bent but very close to being linear for temperatures below 400 K where a linear fit gives an estimation of the barrier height of 127.9 ±1.7 kJ · $mol^{-1}$ close to the FES apparent barrier of ~108 kJ · $mol^{-1}$.

## 4. Conclusions

We have demonstrated a viable strategy for the study of complex reactions. Our strategy relies on the combination of physical insight, enhanced sampling and machine learning. Given the success of this first approach we believe that complex chemical reactions are within the reach of *ab-initio* quality MD simulations. Here we have explicitly treated the case of water driven homogeneous process, but applications to homogeneous and heterogeneous catalytic processes are certainly possible.

## 5. Acknowledgements

The authors thank Prof. Han Wang and Dr. Linfeng Zhang for the helpful discussions, and especially for the support in using and modifying the DP-GEN code and for the valuable help in performing the DeePMD training; Michele Invernizzi, Emanuele Grifoni, Dr. GiovanniMaria Piccini, Dr. Chang Woo Myung and Dr. Valerio Rizzi for useful discussions. M.Y. thanks in particular Linfeng Zhang for support and encouragement. This research was supported by the NCCR MARVEL, funded by the Swiss National Science Foundation and the European Union Grant No. ERC-2014-AdG-670227/VARMET. The calculations were carried out on the Euler cluster of ETH Zurich.

# Supplementary material

## S1 Computational setup

### S1.1 Molecular dynamics setup

The system consists of one molecule of urea and 34 molecules of water within a box with $10 \times 10 \times 10$ Å. AIMD and DeepMD-based simulations were performed using CP2K[1] and LAMMPS[2], respectively. AIMD simulations were performed within the Born-Oppenheimer approximation. In both cases, simulations were carried out at constant volume and temperature with periodic boundary conditions. The temperature was controlled using the stochastic velocity rescaling thermostat[3] with a relaxation time of 0.04 ps, and an integration time step of 0.5 fs was used. The starting configurations of WTMetaD were first equilibrated at the chosen temperatures for at least 30 ps.

### S1.2 DFT Calculations

DFT calculations were calculated using the same setup as Ref.[4]. Namely, PBE exchange-correlation density functional[5] with m-DZVP (2s2p1d/2s1p) Gaussian basis set. Core electrons were treated using the Goedecker−Teter−Hutter (GTH) pseudopotentials[6,7]. These calculations were performed using the Quickstep module of the CP2K program[1]. The single-point energies and forces calculations for the training set used an energy cutoff of 300 Ry. The threshold for energy convergence was set to $10^{-12}$ Hartree and the one related to SCF cycles was set to $10^{-6}$ Hartree.

### S1.3 DeepMD potential

We trained our MetaD-based NN potential using the DeePMD-kit package[8]. In this work, the smooth version of the deep potential model is adopted[9], with a cut-off radius of 6.0 Å. To remove the discontinuity introduced by the cut-off, the $1/r$ term in the network construction is smoothly switched-off by a cosine shape function from 1.0 Å to 6.0 Å. The filter (embedding) network has three layers with (25, 50, 100) nodes/layer and the fitting net is composed of three layers, with 240 nodes each. The network is trained with the ADAM optimizer, with an exponentially decaying learning rate from $1.0 \times 10^{-3}$ to $5.0 \times 10^{-8}$. The batch size was chosen as 4. The pre-factors of the energy and the force terms in the loss function change during the optimization process from 1 to 10 and from 1000 to 1, respectively. The final model used for the production run was trained for $6.0 \times 10^6$ steps (1600 epochs).

### S1.4 Training procedure with DP-GEN

The iterative procedure of the construction of the training set was performed using the DP-GEN package[10]. The criterion used to select new configurations is based on the agreement on the forces predictions made by an ensemble of 4 DPMD potentials, which have been trained on the same reference dataset but with different initial weights. In the following we call model deviation the maximum (over the force components) of the standard deviation on the forces predicted by such an ensemble of models. Whenever the model deviation for one configuration was in the range of [0.2, 0.4] eV/Å the corresponding structure was selected for labeling.

To choose the range for the model deviation, we proceeded in the following way. First, we evaluated the average model deviation on the training set of NN0, and we have set the lower bound to be slightly higher than this value. Choosing a lower bound that is too small, would indeed lead to the selection of structures that are already well-represented based on the ensemble criterion. For the upper bound, we followed the rule of thumb presented in Ref. [10] which suggests using an upper bound for the model deviation that is 0.15 ~ 0.30 eV/Å higher than the lower bound. The purpose of this upper limit is to exclude unphysical configurations from the labeling.

These NN potentials were trained with $1.0 \times 10^6$ training steps. The maximum and the minimum number of configurations to be collected at each iteration (and for each temperature) were chosen to be 200 and 30, respectively. DP-GEN simulations were performed at multiple temperatures (490, 390 and 300K). The active-learning procedure was concluded after 16 iterations.

**S1.5 Metadynamics calculations**

The calculation of the collective variables and the metadynamics biased is done with PLUMED[11]. In the following, we describe the collective variable used and the metadynamics setup. First, we define the coordination numbers which we use below to define the path collective variables. They are calculated as follows:

| CV | Definition | Parameters |
|---|---|---|
| $C_{CN}$ | $C_{CN} = \sum_{i \in N_1, N_2} \dfrac{(1 - \frac{r_{C,i}}{r_0})^m}{(1 - \frac{r_{C,i}}{r_0})^n}$ | $r_{C,i}$: distance between atoms C and i-*th* N; $r_0 = 2.0$ Å; $m = 6$; $n = 12$ |
| $C_{CO}$ | $C_{CO} = \dfrac{(1 - \frac{r_{CO}}{r_0})^m}{(1 - \frac{r_{CO}}{r_0})^n}$ | $r_{CO}$: distance between atoms C and O; $r_0 = 1.8$ Å; $m = 6$; $n = 12$ |
| $C_{N1H}$ | $C_{N1H} = \sum_{i \in H_{all}} \dfrac{(1 - \frac{r_{N1,i}}{r_0})^m}{(1 - \frac{r_{N1,i}}{r_0})^n}$ | $r_{N1,i}$: distance between atoms N1 and i-*th* $H_{all}$; $r_0 = 1.5$ Å; $m = 8$; $n = 16$ |
| $C_{N2H}$ | $C_{N2H} = \sum_{i \in H_{all}} \dfrac{(1 - \frac{r_{N2,i}}{r_0})^m}{(1 - \frac{r_{N2,i}}{r_0})^n}$ | $r_{N2,i}$: distance between atoms N1 and i-*th* $H_{all}$; $r_0 = 1.5$ Å; $m = 8$; $n = 16$ |
| $C_{OH}$ | $C_{OH} = \sum_{i \in H_{all}} \dfrac{(1 - \frac{r_{O,i}}{r_0})^m}{(1 - \frac{r_{O,i}}{r_0})^n}$ | $r_{O,i}$: distance between atoms O and i-*th* $H_{all}$; $r_0 = 1.5$ Å; $m = 8$; $n = 16$ |
| $C_{OHw}$ | $C_{OHw} = \sum_{i \in H_w} \dfrac{(1 - \frac{r_{O,i}}{r_0})^m}{(1 - \frac{r_{O,i}}{r_0})^n}$ | $r_{O,i}$ : distance between atoms O and the $i$-th Hw; $r_0$ = 1.5 Å; $m$ = 8; $n$ = 16 |
| $C_{HOw}$ | $C_{HOw} = \sum_{i \in H} \sum_{j \in O_w} \dfrac{(1 - \frac{r_{ij}}{r_0})^m}{(1 - \frac{r_{ij}}{r_0})^n}$ | $r_{ij}$ : distance between atoms the $i$-th H and the $j$-th Ow; $r_0$ = 1.5 Å, $m$ = 8; $n$ = 16 |

Note: *If it's not specified, the H and O atoms mentioned here those of the solute molecule.*

The definitions of the path CVs $s(\mathbf{d})$ and $z(\mathbf{d})$ are described in the main text. For the construction of such variables, one needs to define two things: the configurational space where to draw the path and the reference configurations. Here we draw inspiration from the urea decomposition pathway used in our previous work [4] and generalize it to include 4 atomic configurations (A1, A3, B1, and B2 in Fig. 3) to define the path in the space of coordination numbers $C_{CN}$, $C_{CO}$ and $C_{N1H} - C_{N2H}$. The parameter λ is chosen to be 0.25. In the metadynamics simulations, the Gaussians adopted have an initial height of 15 kJ/mol and width of 0.1 and 0.2 for the CV $s(\mathbf{d})$ and $z(\mathbf{d})$, respectively. A Gaussian was deposited every 100 fs with a bias factor equal to 20. Once the simulations were performed, we defined new path CVs to distinguish all the pathways and metastable states located by our WTMetaD simulations. In this case, the reference path consisted of six configurations (A1, A3, B1, B2 and the transition states of acid and basic paths (A1 to A3)), adding the $C_{N1H} + C_{N2H}$ variable to the descriptors set.

**S1.6 Kinetic Rates Calculation**

For infrequent metadynamics simulations, bias was added on the two CVs $r_{CN1} - r_{CN2}$ and $C_{N1H} - C_{N2H}$ using Gaussian widths of 0.05 and 0.05, respectively. The combination of these two CVs was able to distinguish the reactant state **A** and product state **B**. To build the static bias we carried out a metadynamics run to fill the reactant basin (state A) up to a preassigned free energy threshold. The threshold was fixed at 90 kJ/mol since the first reactive event during the metadynamics run was detected after depositing 120 kJ/mol. We verified that this value ensured that no bias was deposited on the transition state. This bias was then used as an external potential, and on this biased energy surface, an infrequent metadynamics run was performed.

We calculated the kinetic rates at 12 different temperatures in the range from 300 ~ 510 K (see Table S1). For each temperature, about 100 independent trajectories initiated after thermalization in the **A1** basin were performed, where Gaussians were deposited every 5 ps, with an initial height of 0.4 and 0.6 kJ/mol for T ≥ 370 K and T＜370 K, respectively.

**S1.7 Static calculations**

All static geometries were optimized at the PBE/cc-pVDZ level[5] using the same level of the theory of the *ab-initio* molecular dynamics simulation. The Polarizable Continuum Model (PCM) using the integral equation formalism variant (IEFPCM)[12] was used as an implicit solvation model with water as the solvent. For the TS, intrinsic reaction coordinate (IRC)[13] analysis was performed to verify whether the TS connects the correct minimum structures. All calculations were performed with the Gaussian16 software package[14].

## S2 Additional results

### S2.1 Composition of the training set and accuracy

Our training set was upgraded progressively. The reference configurations are collected from three different sources, which are the initial AIMD simulation (AIMD), the active learning procedure (ACTIVE) and the selection along a collective variable (CV-BASED). Their corresponding composition is reported in Table S1. In this process, we generated 3 main NN potentials that for simplicity will be called NN0 (AIMD), NN1 (AIMD+ACTIVE), and NN2 (AIMD+ACTIVE+CV-BASED).

Table S1. *Composition of the training set. Structures are classified into reactant (A), transition state (TS), and product (B) regions using the coordination number between C and N (Fig. S1).*

| Configuration sets | A | TS | B | Total |
| --- | --- | --- | --- | --- |
| AIMD | 684 | 179 | 687 | 1550 |
| ACTIVE LEARNING | 2991 | 225 | 2581 | 5797 |
| CV-BASED SELECTION | 2064 | 3176 | 1949 | 7189 |
| Cumulative training set (NN2) | 5739 | 3580 | 5217 | 14536 |

More in detail, NN0 was trained including a set of configurations from the available AIMD trajectory of Ref. [4]. This set consisted of 1550 uncorrelated structures that were extracted every 400 steps (0.4 ps) for reactants and products and every 150 steps in the TS region. Including more structures from this AIMD trajectory would mean adding redundant information to the training set with no improvement to the performance of the NN model in terms of stability and accuracy.

NN1 was trained by adding a set of structures collected with an active learning procedure as implemented in DP-GEN where the exploration of the configurational space was enhanced by the use of metadynamics performed at different temperatures. In this step, the configurations are selected only by looking at the deviations of the prediction of the forces of different NN models. This iterative procedure is repeated until no more configurations are selected for labelling.

NN2 contains additional structures extracted from long DP-WTMetaD simulations performed with the final NN1 model at different temperatures, to have a uniform coverage of the reaction pathway. This has been achieved by dividing the configurations into three regions: reactants, products and transition state, according to the value of the coordination number between C and N (as done in Figs. S1 ~ S3). The configurations have been selected from these datasets at even intervals with a shorter stride for the ones in the TS regions to obtain a uniform distribution of configuration between reactants, products and TS regions. We note that to perform this step one needs a reliable model over all the thermodynamically accessible regions, which we achieved with the iterative procedure that led to NN1.

In the following, we report the accuracy tests we performed with the different models. Fig. S1 shows the mean absolute error (MAE) as a function of the C-N coordination number on the training sets of different NN potentials. All the models have been trained with the same parameters for 400 epochs in order to compare them. This allows us to monitor how the error is distributed between reactant (A), products (B) and transition state (TS). From Fig. S1 we observe that by adding new configurations to NN1 we were able to significantly reduce the training error along all the reaction pathway. Note that region A contains the states A1, A2 and A3 and their corresponding transition pathways, B includes the states B1 and B2 (Fig. 3) while TS covers only a limited configurational space related to the highest free energy barrier region. Therefore, it is easier for the NN to model this region when enough configurations are provided in the training set. On the other hand, training the models with a larger number of steps only increases the accuracy uniformly along the reaction pathway (Fig. S2).

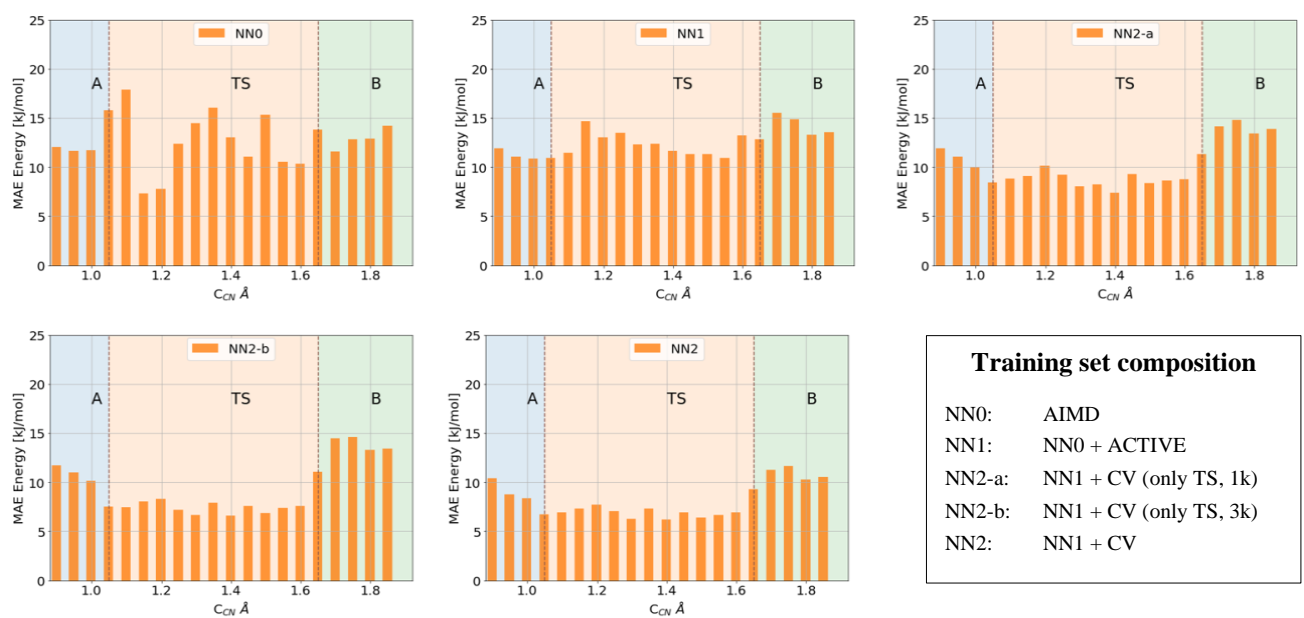

Fig. S1. *MAEs of the energies on the training set for the potentials NN0, NN1, NN2-a, NN2-b and NN2, as a function of the C-N coordination number. The higher the value of the coordination number, the more product-like a configuration is. NN2-a and NN2-b potentials are trained by adding to NN1 only 1000 and 3000 configurations selected from the TS region (without adding structures also from A and B as done in the final potential NN2). All models have been trained for 400 epochs.*

Since the purpose of the DPMD potential is to run MD simulations, a more stringent test is to assess the accuracy of the potential on the generated configurations. In order to do so, we consider for both NN1 and NN2 a test set composed by configurations extracted from DP-WTMetaD simulations performed at multiple temperatures by NN1 and NN2, whose energies and forces were computed at the *ab-initio* level. In Fig. S3 we report the MAE on the test sets for NN1 (panel a) and NN2 (panel b) from which we observe that the addition of the new structures allowed us to obtain a uniform (and lower) error along all the reaction pathway for the structures generated in the DP-WTMetaD simulations. As shown in panel (d) of Fig. S3 the test sets cover all the configurational space of interest.

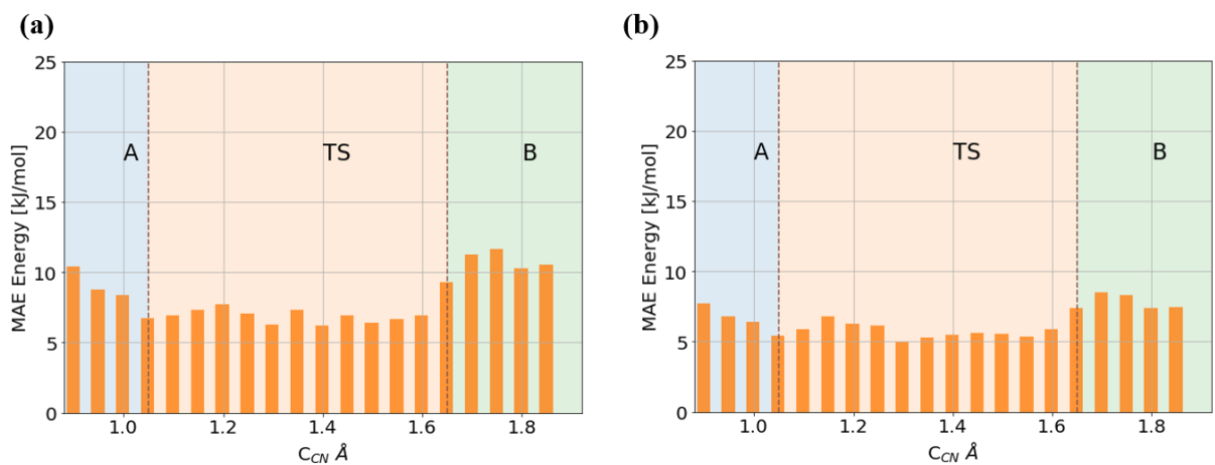

Fig. S2. *MAEs of the energies on the training set for the potentials NN1 and NN2, trained for 400 epochs (a) and for 1600 epochs (b). The latter is the one used for the DP simulations in this work.*

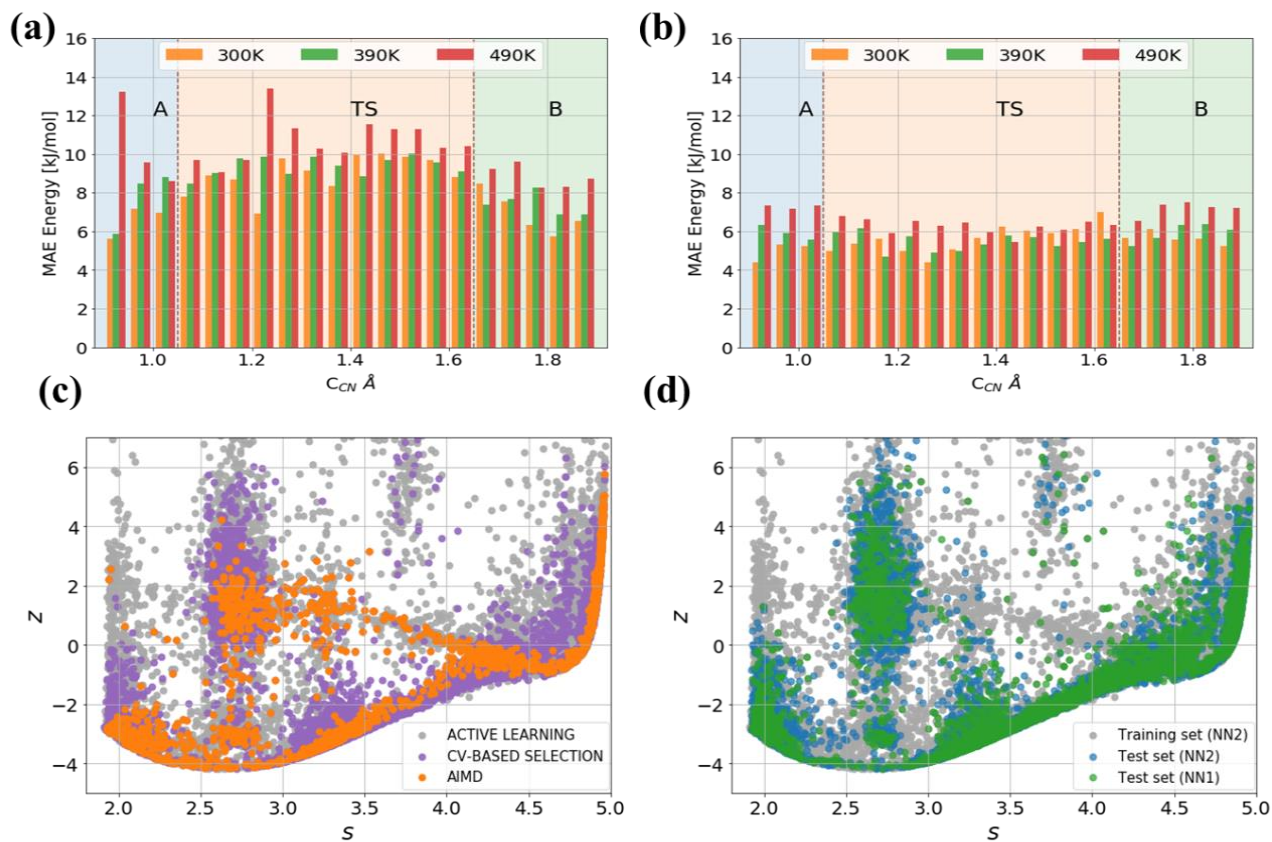

Fig. S3. *MAEs of energies on the test sets as a function of the collective variable for model NN1 (a) and NN2 (b), respectively. The distribution of the training set (c) and test set (d) configurations in the path collective variables space is also reported. The test set of NN1 includes 2442, 2829 and 3709 structures at 300, 390 and 490 K, respectively, while the test set of model NN2 consists of 5919, 5418 and 5534 structures at 300, 390 and 490 K. NN1 and NN2 have been trained for 2000 and 1600 epochs, respectively.*

## S2.2 Validation of the NN potential

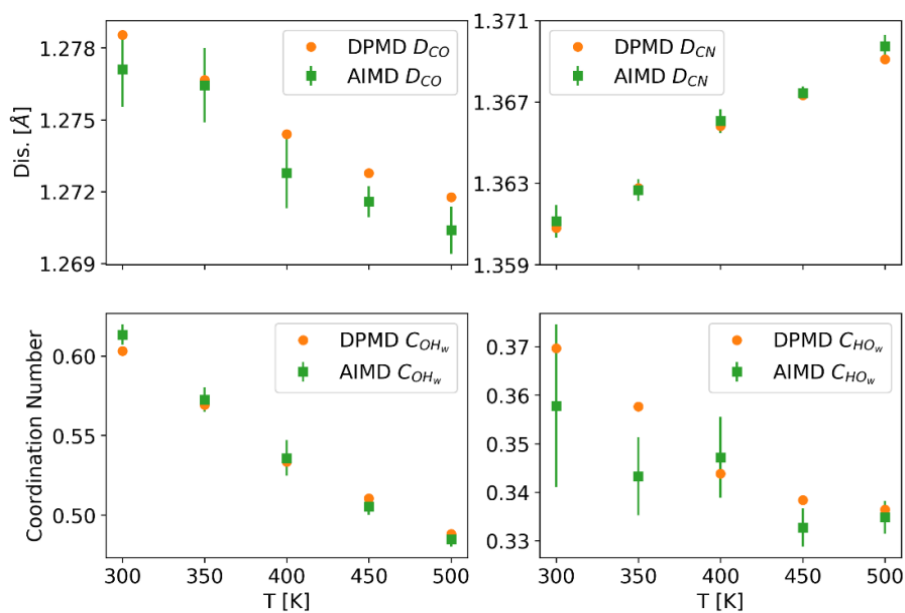

Fig. S4. *Comparison of the conformational distribution of urea in aqueous media at five different temperatures (300, 350, 400, 450 and 500 K) given from DPMD (orange points) and AIMD (green squares). For AIMD calculations, all simulations were initially equilibrated for at least 30 ps after which the data were collected from runs of 380, 300, 100, 100 and 100 ps at 300, 350, 400, 450 and 500 K. The error bars indicate the standard error obtained from AIMD simulations. The standard error from DPMD is within the point markers. $D_{CO}$ and $D_{CN}$ are the (average) distances of bond C-O and C-N in Urea, respectively, $C_{OHw}$ is the coordination number of water hydrogen atoms around the Urea oxygen and $C_{HOw}$ is the coordination number of Urea hydrogen atoms around the water oxygen. In this paper, the standard errors in all figures are calculated from $n=4$ independent simulations, as $\sigma/(n)^{1/2}$, where $\sigma$ is the standard deviation.*

In Fig. S5 ~ S9 we report the comparison of the radial distribution functions for urea at multiple temperatures, between DPMD (solid lines) and AIMD (dashed lines). Shaded areas indicate the standard error obtained from AIMD simulations. Due to the much longer simulation time of DPMD the standard error is negligible.

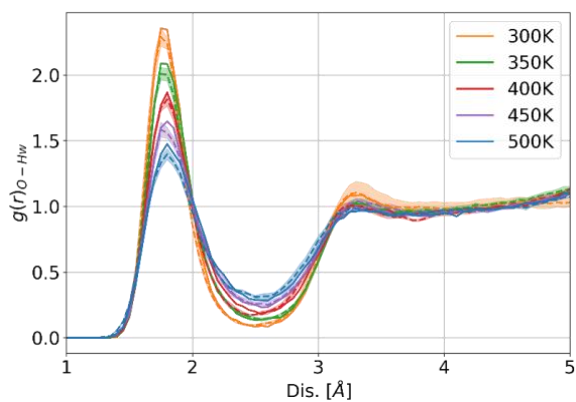
Fig. S5. *O(urea) - H(water)*

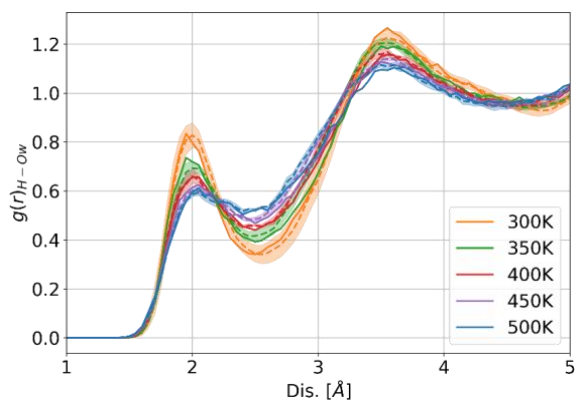
Fig. S6. *H(urea) - O(water)*

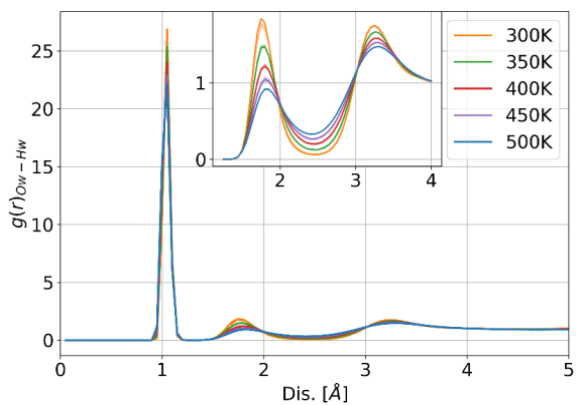

Fig. S7. *O(water)-H(water)*

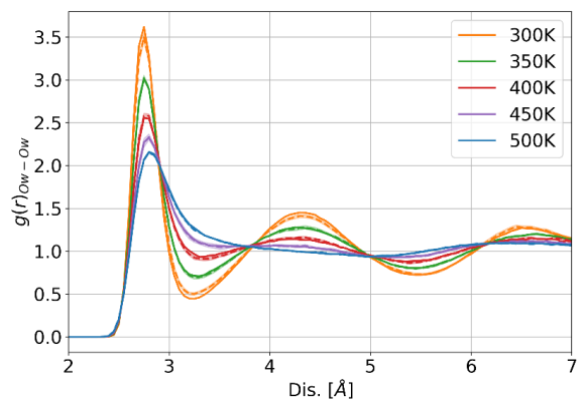

Fig. S8. *O(water)-O(water)*

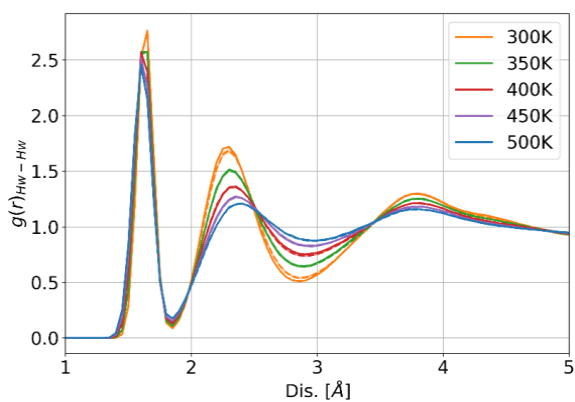

Fig. S9. *H(water)-H(water)*

**S2.3 Convergence of the metadynamics simulations**

In Fig. S10 we monitor the convergence of the free energy difference between reactants and products. For all temperatures the standard deviation is lower than 0.5 $k_BT$. We note that a possible improvement could be the usage of a recently developed evolution of metadynamics, called on-the-fly probability enhanced sampling[15] which has shown to significantly improve the convergence speed and the accuracy of free energy estimates.

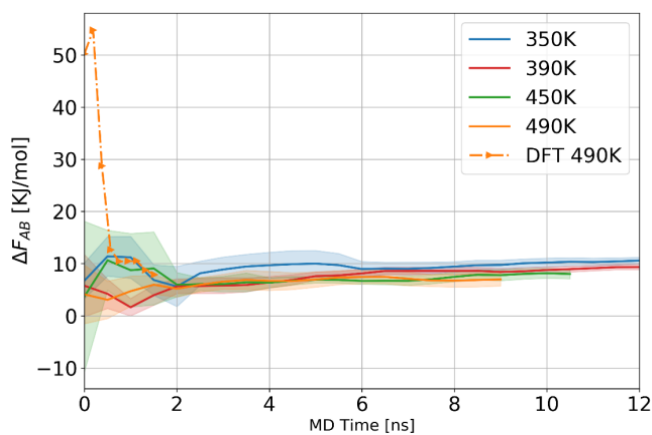

Fig. S10. *Time evolution of the free energy difference between the reactants (A) to products (B), divided by the C-N bond breaking. Shaded areas indicate the standard deviation obtained from 4 independent DP-WTMetaD simulations. AIMD results at 490 K are also illustrated.*

## S2.4 Free energy profiles

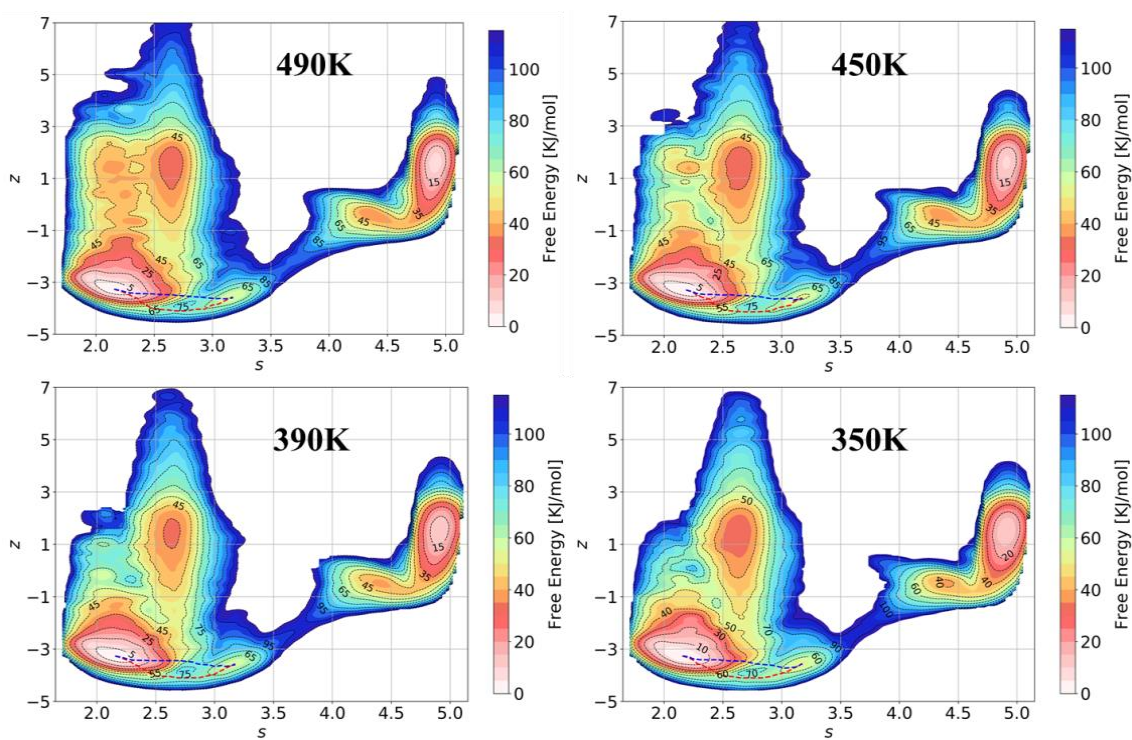

Fig. S11. *Free energy surface of urea decomposition in aqueous solution at multiple temperatures as a function of path collective variables. The dashed lines correspond to the minimum free-energy pathways between A1 and A3. For each temperature, four independent runs of at least 10 ns have been performed.*

## S2.5 Kinetic rates

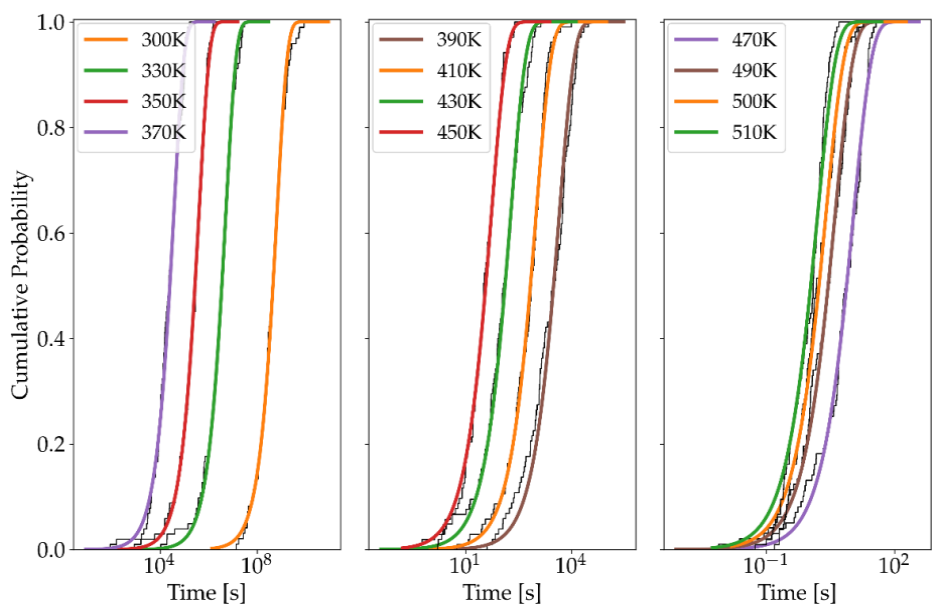

Fig. S12. *Cumulative transition probability (CDF) and relative fit to a Poisson distribution function of the transition times for urea decomposition (from states A1 to B2) at multiple temperatures.*

Table S2. *Average transition time from simulations (μ), theoretical Poisson process (τ), and p-value associated with the KS statistic test (see Ref. [16]) calculated at multiple temperatures.*

| T [K] | μ [s] | τ [s] | K[s$^{-1}$] | Std. Error | *p*-value | N |
|---|---|---|---|---|---|---|
| 300 | 8.02E+08 | 6.37E+08 | 1.57E-09 | 2.12E-11 | 0.88 | 97 |
| 330 | 5.53E+06 | 5.04E+06 | 1.98E-07 | 2.42E-09 | 0.92 | 102 |
| 350 | 3.76E+05 | 3.96E+05 | 2.52E-06 | 2.39E-08 | 0.80 | 102 |
| 370 | 3.80E+04 | 3.68E+04 | 2.72E-05 | 3.60E-07 | 0.71 | 102 |
| 390 | 5.02E+03 | 4.47E+03 | 2.24E-04 | 3.32E-06 | 0.38 | 102 |
| 410 | 9.97E+02 | 9.85E+02 | 1.02E-03 | 1.10E-05 | 0.86 | 104 |
| 430 | 2.14E+02 | 1.88E+02 | 5.32E-03 | 8.37E-05 | 0.67 | 105 |
| 450 | 5.23E+01 | 5.32E+01 | 1.88E-02 | 1.70E-04 | 0.96 | 103 |
| 470 | 1.02E+01 | 1.08E+01 | 9.24E-02 | 8.65E-04 | 0.84 | 99 |
| 490 | 4.04E+00 | 4.01E+00 | 2.49E-01 | 3.47E-03 | 0.65 | 97 |
| 500 | 2.56E+00 | 2.45E+00 | 4.09E-01 | 4.27E-03 | 0.71 | 105 |
| 510 | 1.37E+00 | 1.56E+00 | 6.39E-01 | 5.17E-03 | 0.50 | 105 |

# Supplementary References